\documentclass[11pt,a4paper]{article}
\usepackage{jheppub}

\usepackage{epsfig,multicol,bbm,amsxtra,subfigure,ulem}

\title{ \boldmath Localization of Vector and Fermion Fields on a Thick Brane with a Lump-type Scalar 
Background}

\author[]{Jun-Tong Zhou,}
\author[]{Zhen-Hua Zhao\footnote{Corresponding author}}

\affiliation[]{
    Department of Applied Physics,
    Shandong University of Science and Technology,
    Qingdao, 266590 People's Republic of China}
\emailAdd{Jtzhou@sdust.edu.cn}
\emailAdd{zhaozhh78@sdust.edu.cn}

\abstract{
We study a thick braneworld model in asymptotically anti-de Sitter spacetime, supported by a non-topological lump-type scalar field. Unlike in boson stars, which are also non-topological solutions, the complex scalar field in our setup must have zero oscillation frequency. Consequently, the background scalar field is static and real, which breaks the global $U(1)$ symmetry. This is similar to spontaneous symmetry breaking, but it is not the same. We then investigate the localization of vector and fermion fields within this framework. By introducing a specific non-minimal coupling, we demonstrate that the zero modes of both fields can be successfully localized. Notably, this mechanism allows for the simultaneous localization of both left- and right-handed fermion zero modes.}


\begin{document}
\maketitle

\section{Introduction}\label{sec:intro}
In braneworld theory, the four-dimensional universe is viewed as a brane embedded in a higher-dimensional spacetime. The Arkani-Hamed--Dimopoulos--Dvali (ADD) and Randall--Sundrum-I (RS-I) models addressed the hierarchy problem when the extra dimensions are compact~\cite{arkani-hamed1998,randallLargeMassHierarchy1999}. The Randall--Sundrum-II (RS-II) model subsequently showed that effective four-dimensional Newtonian gravity can be recovered even when the extra dimension is infinite~\cite{randallAlternativeCompactification1999}. By combining these ideas with the domain wall scenario~\cite{rubakov1983}, thick brane models with internal structure were introduced~\cite{dewolfe2000}. Thick branes localize the graviton zero mode and, in an appropriate limit, reduce to the RS-II thin brane~\cite{dewolfe2000,gremm2000b}.

On thick branes, different localization methods apply to the zero modes of different fields. The graviton zero mode is typically localized by analyzing perturbations~\cite{dewolfe2000,gremm2000b,barbosa-cendejas2006,yang2012,moreira2023,tan2024}, whereas vector and fermion fields usually require a non-minimal coupling and a Yukawa coupling, respectively~\cite{kehagias2001,gogberashvili2012,zhao2015,arai2018,koley2005,liu2008,almeida2009,xie2017,li2017,zhu2023}. Most of these methods are based on kink-type background scalar fields, which are topological. Note that the localization of the graviton and scalar zero modes does not depend on whether the background scalar field is topological~\cite{dewolfe2000,gremm2000b,ahmed2014,zhong2018,tan2024,kakushadze2000,bajc2000,kehagias2001}, so the same treatment applies to the non-topological case considered here. In contrast, the localization of vector and fermion fields depends on how they couple to background scalar fields, so their localization properties must be discussed when the background scalar field is non-topological.

In a $(1+1)$-dimensional theory with a single real scalar field, the kink is a topological soliton that connects two minima of the potential. The lump, by contrast, is a non-topological solution whose field configuration leaves one minimum and returns to the same minimum; in the absence of gravity it is unstable under small perturbations~\cite{brihaye2008,bazeia2026,bazeia2026a}. Stable non-topological solutions such as Q-balls~\cite{coleman1985} and boson stars~\cite{rosen1968,kaup1968,ruffini1969,friedberg1976,friedberg1987,lee1992} usually rely on a global $U(1)$ symmetry. In braneworlds, however, gravity may change this situation. Giovannini constructed a lump-type solution on five-dimensional asymptotically anti-de Sitter (AdS) thick branes, showing that non-topological structures do not necessarily require a global $U(1)$ symmetry~\cite{giovannini2006,giovannini2007}. Avelar et al.\ later studied different lump-type solutions in flat spacetime~\cite{avelar2008}. These works focused on constructing lump-type solutions in flat spacetime and in braneworlds, but did not consider how a lump-type background affects the localization of zero modes of various matter fields.

In this paper, we first introduce a complex scalar field and take it to be of harmonic form in order to construct a non-topological solution in the thick brane model. We find that the oscillation frequency of the complex scalar field must vanish, so the background scalar field is static and real. Inspired by the approach of~\cite{avelar2008}, we then construct a piecewise superpotential, obtain an analytic lump-type solution of the background scalar field, and study the localization of vector and fermion zero modes with the lump-type background. In particular, we find a localization mechanism for fermions that is distinct from the usual Yukawa coupling. With a simple non-minimal coupling to the lump-type background, the left- and right-handed fermion zero modes are localized simultaneously, and the localization condition only restricts the range of the dimensionless parameter of the background scalar field, without introducing an additional coupling constant.

The rest of the paper is organized as follows. In Sec.~\ref{sec:model} we present the model and explain why the background scalar field must be a static and real scalar field. In Sec.~\ref{sec:lumps} we use the superpotential method to construct an analytic lump-type solution and discuss the corresponding scalar potential and energy density. In Sec.~\ref{sec:vector} and Sec.~\ref{sec:fermion} we discuss the localization of zero modes of vector and fermion fields. Sec.~\ref{sec:concl} contains our conclusions.

\section{The Model}\label{sec:model}
The line element for the five-dimensional spacetime is given by
\begin{equation}
\mathrm{d} s_5^2 =g_{MN}dx^Mdx^N={e}^{2 A(y)}\eta_{\mu\nu} \mathrm{d}x^\mu\mathrm{d}x^\nu+\mathrm{d}y^2,\label{metric}
\end{equation}
where $M,N,\cdots=\{0,1,2,3,4\}$, $\mu, \nu, \cdots = \{ 0,1,2,3 \}$, and $\eta_{\mu\nu}=\mathrm{diag}{(-1, 1, 1, 1)}$. Here $e^{2A(y)}$ is the warp factor, and the function $A(y)$ depends only on the extra-dimensional coordinate $y$.

In flat spacetime, Derrick's theorem rules out static, localized soliton solutions in pure scalar field theories with more than one spatial dimension~\cite{derrick1964}. The usual ways around this obstruction are to introduce fields with nonzero spin~\cite{rosen1966,friedberg1977} or to construct non-topological solutions by means of time-dependent complex scalar fields~\cite{friedberg1976,friedberg1987,lee1992}. Although the original Derrick theorem is formulated for flat spacetime, time-dependent complex scalar fields also provide a standard ansatz for constructing non-topological solutions in curved spacetime~\cite{kaup1968,friedberg1987,sun2022}.

We introduce a time-dependent complex scalar field $\Phi$ and examine the constraints that the Einstein field equations impose on this ansatz. To this end, we consider an RS-II thick brane model in which gravity is coupled to the complex scalar field $\Phi$, with the action
\begin{equation}
S=\int\mathrm{d}^4x\mathrm{d} y\sqrt{-g}\left(\frac{1}{2\kappa_5^2}R + \mathcal{L}_\Phi \right), \label{Action}
\end{equation}
where $g=\det(g_{MN})$, $R$ is the scalar curvature of the five-dimensional spacetime, and $\kappa_5^2 = 8 \pi G_5$ with $G_5$ the five-dimensional Newton constant. The Lagrangian $\mathcal{L}_\Phi$ is
\begin{equation}
\mathcal{L}_\Phi =- g^{MN}\partial_M\Phi^{*}\partial_N\Phi - U(\Phi^* \Phi), \label{LPhi}
\end{equation}
where $\Phi=\Phi(t,y)$. The
Lagrangian \eqref{LPhi} is invariant under a global $U(1)$ transformation.
In attempts to construct non-topological background solutions, the complex scalar field is often taken in the harmonic form~\cite{kaup1968,ruffini1969,friedberg1976}. Following this practice, we generalize the ansatz to the braneworld
\begin{equation}
\Phi = \frac{1}{\sqrt{2}} e^{ - i \omega t} \phi(y),\label{tphi}
\end{equation}
where $\phi(y)$ is a real scalar field and $\omega$ is a constant.

By varying the action \eqref{Action}, we obtain the Einstein equations
\begin{equation}
R_{MN} - \frac{1}{2} g_{MN} R =\kappa_5^2 T_{MN},
\end{equation}
where
\begin{equation}
T_{MN} \equiv -\frac{2}{\sqrt{-g}}\frac{\delta\left(\sqrt{-g}\,\mathcal{L}_{\Phi}\right)}{\delta g^{MN}} =\partial_{M}\Phi^*\,\partial_{N}\Phi+\partial_{N}\Phi^*\,\partial_{M}\Phi+ g_{MN}\mathcal{L}_{\Phi}.\label{T}
\end{equation}
Substituting Eq.~\eqref{tphi} into the action \eqref{Action}, we obtain the components of the Einstein field equations
\begin{align}
&(0,0): \quad -\kappa_5^2\omega ^2 \phi (y)^2 - e^{2 A(y)} \left[12 A'(y)^2+\kappa_5^2\phi '(y)^2+ 2\kappa_5^2U(\phi)+6 A''(y)\right]=0, \label{eq01} \\
\notag\\
&(i,i): \quad -\kappa_5^2\omega ^2 \phi (y)^2 +e^{2 A(y)} \left[12 A'(y)^2 + \kappa_5^2\phi '(y)^2 + 2\kappa_5^2U(\phi) + 6 A''(y)\right]=0, \label{eq11}\\
\notag\\
&(4,4): \quad -\kappa_5^2 \omega ^2 \phi (y)^2+e^{2A(y)}\left[12 A'(y)^2-\kappa_5^2\phi '(y)^2+2\kappa_5^2U(\phi)\right] =0, \label{eq21}
\end{align}
where the prime denotes the derivative with respect to $y$, and $i=1,2,3$. Adding Eqs.~\eqref{eq01} and \eqref{eq11} gives
$$-2 \kappa_5^2\omega^2 \phi(y)^2 =0,$$
which must hold for any non-trivial scalar field $\phi(y)$, hence
\begin{align}
\omega = 0.
\end{align}
Thus, in our model, the Einstein field equations require $\omega = 0$, so that the background scalar field takes the static, real form 
\begin{align}
\Phi=\frac{\phi}{\sqrt{2}}. \label{SSBl}
\end{align}

Note that the phase of $\Phi$ is fixed by the field equations rather than selected by the scalar potential, so this is not exactly spontaneous symmetry breaking (SSB). Nevertheless, for the real background scalar field \eqref{SSBl}, the global $U(1)$ symmetry is broken, which resembles SSB. We therefore refer to it as an SSB-like breaking of the global $U(1)$ symmetry.

For $\omega=0$, the Einstein field equations \eqref{eq01}--\eqref{eq21} reduce to
\begin{align}
&3 A''(y) +\kappa_5^2\phi '(y)^2 = 0, \label{eq0121}\\
\notag\\
&12 A'(y)^2-\kappa_5^2\phi '(y)^2+2\kappa_5^2U(\phi) = 0, \label{eq2121}
\end{align}
and the Lagrangian \eqref{LPhi} of the complex scalar field becomes that of a real scalar field
\begin{align}
\mathcal{L}_\Phi =\mathcal{L}_\phi=- \frac{1}{2} g^{MN}\partial_M\phi \partial_N\phi - U(\phi). \label{Lphi}
\end{align}
The equation of motion for $\phi(y)$ is
\begin{equation}
\partial_y^2 \phi + 4 \partial_y A \partial_y \phi - \frac{ \partial U(\phi)}{\partial\phi} =0. \label{eq5}
\end{equation}
In the following, we look for lump-type solutions satisfying Eqs.~\eqref{eq0121}, \eqref{eq2121} and \eqref{eq5}.

\section{Lump-type Solution}\label{sec3}\label{sec:lumps}
To obtain lump-type solutions on the five-dimensional RS-II thick brane, we employ the superpotential method~\cite{cvetic1992,dewolfe2000,skenderis1999,gremm2000b,avelar2008} and introduce a superpotential $W(\phi)$ satisfying
\begin{equation}
\partial_y A=-\frac{1}{3}W,\quad\partial_y\phi=\frac{W_\phi}{\kappa_5^2}, \label{SP}
\end{equation}
where $W_\phi \equiv dW/d\phi$. With $U(\phi)$ of the form
\begin{equation}
U(\phi)= \frac{1}{2\kappa_5^{4}} \left( \frac{dW}{d\phi} \right)^2 - \frac{2}{3\kappa_5^{2}} W^2, \label{U}
\end{equation}
a suitable superpotential yields $A(y)$ and $\phi(y)$ that solve the field equations \eqref{eq0121}, \eqref{eq2121}, and \eqref{eq5}. 

Since we are constructing lump-type solutions, $\phi(y)$ must be an even function of $y$. If $W$ is a single-valued function of $\phi$, however, the direction in which $\phi$ evolves depends only on $\phi$ itself, so $\phi(y)$ varies monotonically with $y$ and cannot be even. We therefore need a superpotential with a piecewise structure
\begin{equation}
W(\phi)=
\begin{cases}
\displaystyle +\frac{c_2\left(c_1^2-\kappa_5^2\phi^2\right)^{3/2}}{3c_1},& y>0,\\[6pt]
\displaystyle -\frac{c_2\left(c_1^2-\kappa_5^2\phi^2\right)^{3/2}}{3c_1},& y<0. 
\end{cases}\label{W}
\end{equation}
Here $c_1$ and $c_2$ are parameters of the superpotential. Eqs.~\eqref{SP} and \eqref{W} require $c_1$ to be dimensionless and $c_2$ to have dimension $[c_2]=[L]^{-1}$. The first-order relations \eqref{SP} hold separately for $y>0$ and $y<0$, with $W(\phi)$ taking the two branches of opposite sign. Since a smooth thick brane solution requires $A'(y)$ to be continuous, it follows from \eqref{SP} that $W(\phi(y))$ must be continuous in $y$, so that at $y=0$ the limits of the two branches in \eqref{W} must both vanish, namely $W(\phi(0))=0$, which gives $\phi(0)=c_1/\kappa_5$ and is consistent with $c_1$ being dimensionless.

\begin{figure}[htbp]
    \centering
    \subfigure[]{
        \includegraphics[width=0.39\textwidth]{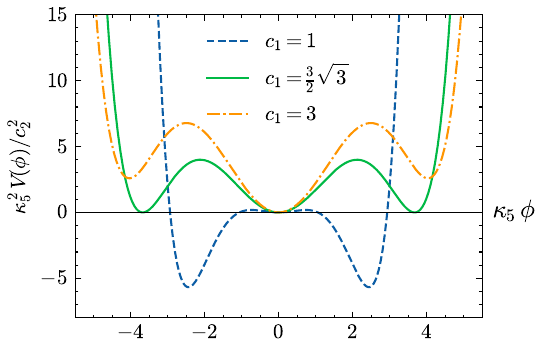}
        \label{f1}
        }
    \subfigure[]{
        \includegraphics[width=0.39\textwidth]{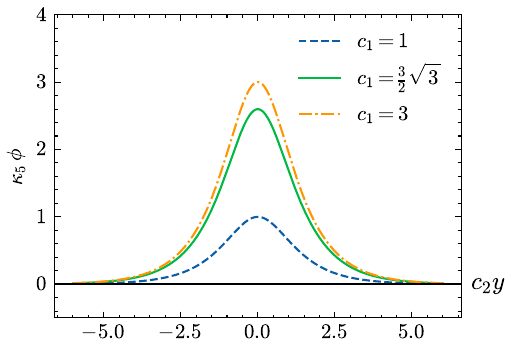}
        \label{f2}
        }   
    \caption{The lump-type solution and the scalar potential, where \ref{f1} shows $\kappa_5^2 V/c_2^2$ versus $\kappa_5\phi$ and \ref{f2} shows $\kappa_5\,\phi$ versus $c_2\,y$. The quantities $\kappa_5^2 V/c_2^2$, $\kappa_5\phi$, and $c_2y$ are all dimensionless.}
    \label{fig:1_figures}
\end{figure}

Substituting the piecewise superpotential \eqref{W} into the potential \eqref{U}, we find
\begin{equation}
U(\phi) = -\frac{2 c_1^4 c_2^2 }{27\kappa_5^{2}} +\phi^2 \left(\frac{2 c_1^2 c_2^2 }{9}+\frac{c_2^2}{2}\right)-\phi^4 \left(\frac{c_2^2\kappa_5^{2}}{2 c_1^2}+\frac{2 c_2^2 \kappa_5^{2}}{9}\right)+\phi^6\frac{2 c_2^2 \kappa_5^{4}}{27 c_1^2}.
\end{equation}
We split the potential into a scalar potential $V(\phi)$ and a cosmological constant $\Lambda_5$
\begin{equation}
U(\phi)=V(\phi)+\Lambda_5,
\end{equation}
where
\begin{equation}
V(\phi)=
\frac{c_2^2}{18c_1^2}
\left[
(4c_1^2+9)\phi^2
\left(c_1^2-\kappa_5^2\phi^2\right)
+\frac{4}{3}\kappa_5^4\phi^6
\right].\label{V}
\end{equation}
\begin{equation}
\Lambda_5=  -\frac{2 c_1^4 c_2^2}{27\kappa_5^{2}}. \label{lamb}
\end{equation}
The potential \eqref{V} is similar in form to that obtained by Wan et al.\ in six-dimensional spacetime~\cite{wan2021}. The difference is that we have restored $\kappa_5$. For $\kappa_5=1$, the coefficients of the terms involving $9\phi^2
\left(c_1^2-\kappa_5^2\phi^2\right)$ in Eq.~\eqref{V} coincide with those of Wan et al., while all other coefficients are scaled by a factor $16/15$. As shown in Fig.~\ref{f1}, for $c_1=\frac{3\sqrt{3}}{2}$ the quantity $\kappa_5^2 V/c_2^2$ has minima at $\kappa_5\phi=0$ and $\kappa_5\phi=\pm \frac{3\sqrt{6}}{2}$. Since $\phi=0$ is a local minimum of $V$ for any $c_1>0$, to make it the global minimum of the scalar potential, we take $c_1\geq \frac{3\sqrt{3}}{2}$.

With the additional choice $A(0)=0$, we obtain the analytic solution
\begin{align}
A(y)&= -\frac{1}{18} c_1^2 \left(2 \log(\cosh(c_2 y)) + \operatorname{sech}^2(c_2 y) -1\right),\label{solutionA}\\ 
 \phi(y) &=  \frac{c_1}{\kappa_5} \  \mathrm{sech} ( c_2 y). \label{solution}
\end{align} 
Unlike kink-type solutions~\cite{dewolfe2000,gremm2000b,kehagias2001}, the lump-type solution satisfies $\phi(y\rightarrow\pm\infty)\rightarrow 0$, as shown in Fig.~\ref{f2}.

As $y\to\pm\infty$, $V(\phi)\to 0$, while the RS-II model requires $A'\rightarrow\mp k= \mp \frac{1}{9}c_1^2c_2$ with $k>0$, which implies
\begin{align}
c_2>0.\label{c2}
\end{align}
From the expression for $k$ we have $c_2=\frac{9k}{c_1^2}$. Substituting this into Eq.~\eqref{lamb} gives $\Lambda_5=-6k^2/\kappa_5^2$. Following the RS-II convention, we write the coefficient of $R$ in the action \eqref{Action} as $1/(2\kappa_5^2)=2M_5^3$, or equivalently $2\kappa_5^2=\frac{1}{2M_5^3}$, with $M_5$ the five-dimensional Planck mass~\cite{randallAlternativeCompactification1999}. The cosmological constant can then be written as $\Lambda_5=-24M_5^3k^2$, which coincides with the vacuum cosmological constant of the RS-II model~\cite{randallAlternativeCompactification1999}.

\begin{figure}[htbp]
    \centering
    \subfigure[]{
        \includegraphics[width=0.39\textwidth]{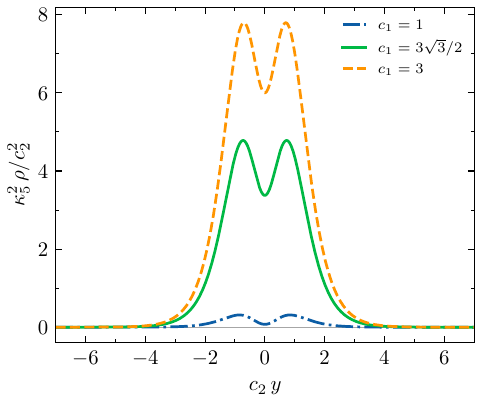}
        \label{f3}
        }
    \subfigure[]{
        \includegraphics[width=0.39\textwidth]{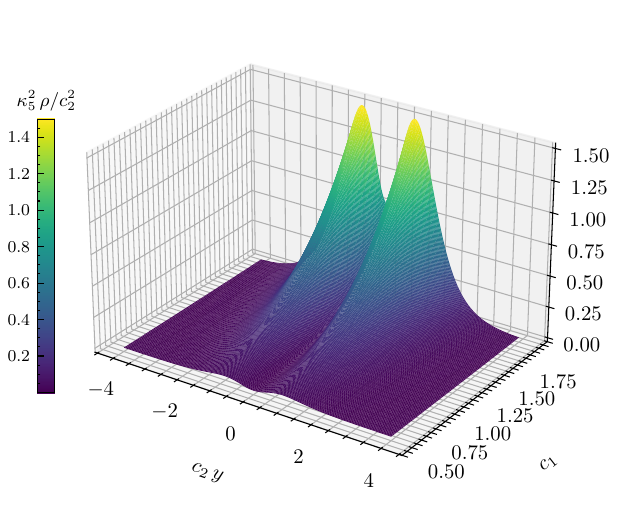}
        \label{f5}
    }
    \caption{Energy density $\kappa_5^2\,\rho/c_2^2$ versus $c_2y$ for different parameter values, where $\kappa_5^2 \rho/c_2^2$ is dimensionless. \ref{f3} shows the double-peak structure for a particular value of $c_1$, and \ref{f5} shows how the energy density profile of the background scalar field changes with $c_1$.}
    \label{fig:2_figures}
\end{figure}

The energy density of the background scalar field is $\rho=T_{MN} u^M u^N$, where $u^M$ is the five-velocity satisfying $g_{MN}u^Mu^N=-1$. For a static observer, $\rho=T_{00} u^0 u^0$. Here, $\rho$ is the energy density of the background scalar field with the constant $\Lambda_5$ subtracted. The internal structure of the thick brane is shown in Fig.~\ref{fig:2_figures}. As Fig.~\ref{f5} shows, the energy density grows markedly with increasing $c_1$, most noticeably near $y=0$ and around the double-peak maxima.

\section{Localization of Vector Fields}\label{sec:vector}
A vector field minimally coupled to background scalar fields cannot be localized in RS-II braneworlds~\cite{pomarol2000}, so a suitable non-minimal coupling is usually required. Chumbes introduced the following mechanism~\cite{chumbes2012}:
\begin{equation}
S=-\frac{1}{4}\int d^5x \sqrt{-g}G(\phi) \mathcal{F}_{MN}\mathcal{F}^{MN}.
\end{equation}
There, $\phi$ is a kink-type solution, and $G(\phi)$ is required to be finite at the center of the brane and to vanish, together with its derivative, at infinity~\cite{chumbes2012}. A kink-type background scalar field, however, approaches a nonzero constant at infinity and cannot play the role of $G(\phi)$ by itself. The lump-type solution \eqref{solution} considered here, by contrast, is an even function that decays to zero asymptotically and can therefore serve directly as the coupling function. The action then reads
\begin{equation}
S=-\frac{1}{4}\int d^5x \sqrt{-g}\,\kappa_5\phi(y) \mathcal{F}_{MN}\mathcal{F}^{MN},\label{gauge}
\end{equation}
where the field strength is $\mathcal{F}_{MN}=\partial_M \mathcal{A}_N - \partial_N \mathcal{A}_M$.

The coupling function multiplying the kinetic term of the vector field is $\kappa_5\phi(y)$. Since the lump-type background satisfies $\phi(y)\to0$ as $y\to\pm\infty$, this coupling function vanishes asymptotically. This is similar to the dielectric structure through which the Friedberg--Lee bag model realizes color confinement~\cite{friedberg1977,friedberg1977a,friedberg1978,goldflam1982,haider1993}. In that model, hadrons are non-topological solitons and the dielectric constant $K(\sigma)$ multiplies the gluon kinetic term, turning it into $-\frac{1}{4}K(\sigma)F^a_{\mu\nu}F^{a\mu\nu}$. With $K(\sigma)=1$ at $\sigma=0$ and $K(\sigma)\to 0$ as $\sigma$ approaches its asymptotic value, the color electric field is confined inside the bag. In our action \eqref{gauge}, the dimensionless coupling function $\kappa_5\phi(y)$ acts as a dielectric constant.

Substituting the metric \eqref{metric} into the action \eqref{gauge} and, for convenience, fixing the gauge
\begin{equation}
\mathcal{A}_4=0, \qquad \partial^\mu \mathcal{A}_\mu=0,
\end{equation}
and using the Kaluza--Klein (KK) decomposition 
\begin{align}
\mathcal{A}_\mu = \sum_n a_\mu^{(n)}(x)\beta_n(y),
\end{align}
we find that the massless mode $\beta_0(y)$ satisfies
\begin{equation}
\beta_0'' + \left(\frac{\phi'}{\phi} + 2A'\right)\beta_0' =0,\label{ve}
\end{equation}
and its action reads
\begin{equation}
S = -\frac{1}{4}\int \mathrm{d}y\,\kappa_5\phi(y)\beta_0^2(y) \int \mathrm{d}^4 x\bigl(f^{(0)}_{\mu\nu} f^{(0)\mu\nu}),
\end{equation}
where $f^{(n)}_{\mu\nu} = \partial_\mu a_\nu^{(n)} - \partial_\nu a_\mu^{(n)}$ is the four-dimensional field strength tensor. Localization of the vector field zero mode requires
\begin{equation}
I \equiv \int_{-\infty}^{+\infty} \mathrm{d}y\,\kappa_5\phi(y)\beta_0^2(y)<\infty.\label{con4}
\end{equation}

We now write
\begin{equation}
\beta_0=e^{-\lambda(y)}\hat{\beta}_0(y),
\end{equation}
with $2\lambda' = 2A' +\frac{\phi'}{\phi}$, so that Eq.~\eqref{ve} becomes
\begin{equation}
-\hat{\beta}_0'' + (\lambda''+\lambda'^2)\hat{\beta}_0 = 0.
\end{equation}
This equation is solved by
\begin{align}
\hat{\beta}_0 &= q_1e^{\lambda(y)}, 
\end{align}
namely
\begin{align}
\beta_0&=q_1,\label{gaugesolution1}
\end{align}
with $q_1$ an arbitrary constant. Substituting Eqs.~\eqref{solution} and \eqref{gaugesolution1} into the condition \eqref{con4} gives
\begin{equation}
q_1^2 \int_{-\infty}^{+\infty} dy {c_1}\mathrm{sech} ( c_2 y)<\infty.
\end{equation}
The localization condition is therefore satisfied. Like the graviton~\cite{dewolfe2000,gremm2000b,kakushadze2000} and scalar zero modes~\cite{bajc2000}, this zero mode has a constant profile \eqref{gaugesolution1}; unlike them, however, it is localized not through the warp factor but through its coupling to the lump-type background.

\section{Localization of Fermion Fields}\label{sec:fermion}
In RS-II braneworlds, free fermions are generically not localized on the brane~\cite{randallAlternativeCompactification1999,bajc2000}. Localization is typically achieved by a coupling between fermions and the background scalar field~\cite{ringeval2002,melfo2006,liang2009,chumbes2011,barbosa-cendejas2015}, usually a Yukawa coupling to a kink-type scalar field. Since the solution of the background scalar field \eqref{solution} considered here is a lump rather than a kink, the Yukawa coupling cannot localize fermions. Motivated by the vector field case, we introduce a new coupling between a spin-$\frac{1}{2}$ fermion $\Psi$ and the lump-type background. The corresponding five-dimensional fermion action is
\begin{equation}
S_{1/2}=\int d^{5}x\,\sqrt{-g}\,
\left(\kappa_5\phi\,\bar\Psi\,\Gamma^{M}D_{M}\Psi\right),\label{s12}
\end{equation}
where $\Gamma^{M}$ are the gamma matrices in curved spacetime, $D_{M}\Psi=(\partial_M+\hat{\omega}_M)\Psi$, and $\hat{\omega}_M$ is the spin connection. In our setup, with the metric \eqref{metric}, we have $\Gamma^{M}=(e^{-A}\gamma^{\mu},\gamma^{5})$, and
\begin{equation}
\Gamma^M D_M=e^{-A}\gamma^\mu\partial_\mu+\gamma^5\big(\partial_y+2A'\big).
\label{gammad}
\end{equation}

Redefining $\bar{\Psi}$ and $\Psi$ according to
\begin{align}
\bar{\Psi} \rightarrow \phi^{-\frac{1}{2}}\bar{\tilde\Psi}, \quad
\Psi \rightarrow \phi^{-\frac{1}{2}}{\tilde\Psi}, \label{redefine}
\end{align}
and substituting into the action \eqref{s12} yields
\begin{equation}
S_{1/2}=\int d^5x\,\sqrt{-g}\;\kappa_5\bar{\tilde\Psi}\Big[
\Gamma^M D_M-\frac{\phi'}{2\phi}\gamma^5\Big]\tilde\Psi .
\end{equation}
Varying with respect to $\bar{\tilde\Psi}$ gives
\begin{equation}
\Big[\Gamma^M D_M-\frac{\phi'}{2\phi}\gamma^5\Big]\tilde\Psi=0 .\label{fermioneq1}
\end{equation}
Substituting Eq.~\eqref{gammad} into Eq.~\eqref{fermioneq1} gives
\begin{equation}
{\;e^{-A}\gamma^\mu\partial_\mu\tilde\Psi
+\gamma^5\Big(\partial_y+2A'-\frac{\phi'}{2\phi}\Big)\tilde\Psi=0.}
\label{eq:eom}
\end{equation}

Using the usual KK and chiral decomposition
\begin{equation}
\tilde{\Psi}=\sum_{n}(\tilde{\psi}_{L}^{(n)}f_{L}^{(n)}+\tilde{\psi}_{R}^{(n)}f_{R}^{(n)})  ,\label{fermionkk}
\end{equation}
we obtain
\begin{equation}
e^{-A}\gamma^\mu\partial_\mu\big(f_L^{(n)}\tilde{\psi}_L^{(n)}+f_R^{(n)}\tilde{\psi}_R^{(n)}\big)
+\gamma^5 (\partial_y+2A'-\frac{\phi'}{2\phi})\big(f_L^{(n)}\tilde{\psi}_L^{(n)}+f_R^{(n)}\tilde{\psi}_R^{(n)}\big)=0 .
\label{fermioneq3}
\end{equation}
With the chirality properties
\begin{equation}
\gamma^5\tilde{\psi}_L^{(n)}=-\tilde{\psi}_L^{(n)},\qquad \gamma^5\tilde{\psi}_R^{(n)}=+\tilde{\psi}_R^{(n)},
\label{eq:chiral}
\end{equation}
and the four-dimensional Dirac equations
\begin{equation}
\gamma^\mu\partial_\mu\tilde{\psi}_L^{(n)}=m_n\tilde{\psi}_R^{(n)},\qquad
\gamma^\mu\partial_\mu\tilde{\psi}_R^{(n)}=m_n\tilde{\psi}_L^{(n)},
\label{eq:dirac4}
\end{equation}
we take the zero mode $m_0=0$ and find
\begin{equation}
\Big(\partial_y+2A'-\frac{\phi'}{2\phi}\Big)f_L^{(0)}
=\Big(\partial_y+2A'-\frac{\phi'}{2\phi}\Big)f_R^{(0)}=0 .
\label{eq:fsame}
\end{equation}
The left- and right-handed zero modes therefore obey identical equations and must have the same functional form.

With the KK decomposition \eqref{fermionkk}, the zero mode action takes the form
\begin{equation}
S_{1/2}^{(0)} = 
\int d^4x \, \bar{\tilde{\psi}}_{L}^{(0)}\gamma^\mu\partial_\mu \tilde{\psi}_{L}^{(0)} \cdot I_L
+ \int d^4x \, \bar{\tilde{\psi}}_{R}^{(0)}\gamma^\mu\partial_\mu\tilde{\psi}_{R}^{(0)} \cdot I_R,
\label{eq:zero-mode-action}
\end{equation}
where 
\begin{equation}
I_L = \kappa_5\int_{-\infty}^{\infty} dy\, e^{3A} (f_{L}^{(0)})\,^2, \qquad
I_R =\kappa_5 \int_{-\infty}^{\infty} dy\, e^{3A} (f_{R}^{(0)})\,^2.
\label{eq:IL-IR-def}
\end{equation}
Solving Eq.~\eqref{eq:fsame}, we obtain
\begin{align}
f_{L,R}(y)=C_{L,R} e^{-2A(y)} \sqrt{\phi(y)}.\label{fl}
\end{align}
Since the equations for the left- and right-handed zero modes differ only in the constants, their solutions share the same form. Localization requires $I_{L,R}<\infty$, and from Eqs.~\eqref{eq:IL-IR-def} and \eqref{fl} we have
\begin{equation}
I_{L,R}=C_{L,R}^2\, \kappa_5  \int_{-\infty}^{\infty}
  e^{-A}\phi dy. \label{lc3}
\end{equation}
Substituting the solutions for $A$ \eqref{solutionA} and $\phi$ \eqref{solution}, we obtain the localization condition 
\begin{equation}
C_{L,R}^2  \int_{-\infty}^{\infty} {c_1 \exp\left(\frac{c_1^2}{18} \left(\operatorname{sech}^2(c_2 y) - 1\right)\right) \cosh^{\frac{c_1^2}{9} - 1}(c_2 y)}dy<\infty.\label{fermionlc}
\end{equation}
When $y\to\pm\infty$, 
\begin{align}
e^{-A(y)}\phi(y)
\to
\frac{c_1}{\kappa_5}
2^{\left(1-\frac{c_1^2}{9}\right)}
e^{-\frac{c_1^2}{18}}
e^{\left(\frac{c_1^2}{9}-1\right)c_2|y|}.
\end{align}
So the integrand is asymptotically dominated by $e^{\left(\frac{c_1^2}{9}-1\right)c_2|y|}$. The integral in Eq.~\eqref{lc3} therefore converges provided $\left(\frac{c_1^2}{9}-1\right)c_2<0$. From Eq.~\eqref{c2}, $c_2>0$, and therefore the fermion zero modes are localized when
\begin{align}
0<c_1<3.
\end{align}

The left- and right-handed zero modes have wave functions of identical form and differ only by constants, so their localization conditions are identical as well. In this mechanism, the two chiralities are therefore either both localized or both non-localized. A mixed situation in which only one chirality is localized cannot occur.

The constraint on $c_1$ is worth emphasizing. From Eqs.~\eqref{solutionA} and \eqref{solution}, this dimensionless parameter appears in both the warp factor $e^{2A}$ and the scalar field $\phi$. For fixed $c_2$, the restriction $0<c_1<3$ has a clear physical meaning. As Fig.~\ref{fig:2_figures} shows, the peak of the energy density of the background scalar field grows with $c_1$, so limiting $c_1$ amounts to bounding that peak. In other words, once the peak of the energy density becomes too high, the fermion zero modes can no longer be localized on the brane.

\section{Conclusions and Discussion}\label{Cons}\label{sec:concl}
In this paper we studied a thick braneworld model in asymptotically AdS spacetime supported by a lump-type scalar field and showed how the zero modes of vector and fermion fields can be localized in this framework.

We first discussed the complex scalar field in the harmonic form, and found that the Einstein field equations force the oscillation frequency to be zero, so the background scalar field must be static and real. Since the field equations constrain the field to be real, the global $U(1)$ symmetry is broken. This mechanism is similar to but different from spontaneous symmetry breaking, and we called it SSB-like breaking.

We then constructed a lump-type solution of the background scalar field with a piecewise superpotential. Its profile is even in the extra dimension, reaches a maximum at the brane center, and decays to zero at infinity, giving rise to a double-peak energy density structure. The shape of the scalar potential is determined by the parameter $c_1$. For $c_1>3\sqrt{3}/2$, the point $\phi=0$ is the unique global minimum of the potential, whereas at $c_1=3\sqrt{3}/2$ it is degenerate with two additional global minima located at nonzero field values.

Next, in order to localize the vector and fermion fields, we introduced a non-minimal coupling, in which the kinetic term of each field is multiplied directly by the background scalar field. It should be emphasized that, for the fermion, this coupling localizes the left- and right-handed zero modes on the brane simultaneously, in contrast to the Yukawa coupling, for which the two chiral zero modes cannot be localized at the same time.

Finally, requiring $\phi=0$ to be a global minimum of the scalar potential restricts the allowed range to $3\sqrt{3}/2\le c_1<3$. For smaller $c_1$ the global minimum moves away from $\phi=0$, which would then be only a local minimum of the potential. For larger $c_1$, the fermion zero modes can no longer be localized. Since the peak of the energy density of the background scalar field grows with $c_1$, the allowed range of $c_1$ can also be interpreted as a bound on the height of this peak.

\bibliographystyle{JHEP}
\bibliography{ref}

\end{document}